\documentclass[10pt,journal,letterpaper]{IEEEtran}

\ifCLASSOPTIONcompsoc

\else

\fi

\ifCLASSINFOpdf

\else

\fi

\usepackage{amsmath}
\usepackage{amsthm}
\usepackage{times}
\usepackage{algorithm}
\usepackage[noend]{algorithmic}
\usepackage{algorithmic}
\usepackage{graphicx}
\usepackage{verbatim}
\usepackage{amssymb}
\usepackage{cite}
\usepackage{epsfig}
\usepackage{color}
\usepackage{url}
\usepackage{xcolor,colortbl}
\usepackage{multirow}
\usepackage{grffile}
\usepackage{epstopdf}
\usepackage{soul}

\begin{document}
\title{Semi-supervised Deep Reinforcement Learning in Support of IoT and Smart City  Services}

\author{Mehdi Mohammadi,~\IEEEmembership{Graduate Student Member,~IEEE,} Ala Al-Fuqaha,~\IEEEmembership{Senior Member,~IEEE,} Mohsen Guizani,~\IEEEmembership{Fellow,~IEEE,} and~Jun-Seok Oh
\IEEEcompsocitemizethanks{\IEEEcompsocthanksitem 
This publication was made possible by NPRP grant\# [7‑1113-1-199] from the Qatar National Research Fund (a member of Qatar Foundation). The statements made herein are solely the responsibility of the authors.

Mehdi Mohammadi and Ala Al-Fuqaha are with the Department of Computer Science, Western Michigan University, Kalamazoo, MI 49008 USA (e-mail: \{mehdi.mohammadi, ala.al-fuqaha\}@wmich.edu). Mohsen Guizani is with the department of Electrical and Computer Engineering, University of Idaho, Moscow, ID 83844 USA (e-mail: mguizani@ieee.org). Jun-Seok Oh is with the Department of Civil and Construction Engineering, Western Michigan University, Kalamazoo, MI 49008 USA (e-mail: jun.oh@wmich.edu).

}

}

\markboth{IEEE Internet of Things Journal,~Vol.~x, No.~x, xxxxx~2017}%
{Shell \MakeLowercase{\textit{et al.}}: Bare Demo of IEEEtran.cls for Computer Society Journals}

\IEEEpubid{\begin{minipage}{\textwidth}\ \\[30pt] \centering
  Copyright (c) 2017 IEEE. Personal use of this material is permitted. However, permission to use this material for any other purposes must be obtained from the IEEE by sending a request to pubs-permissions@ieee.org.
\end{minipage}} 

\IEEEcompsoctitleabstractindextext{
\begin{abstract}

Smart services are an important element of the smart cities and the Internet of Things (IoT) ecosystems where the intelligence behind the services is obtained and improved through the sensory data. Providing a large amount of training data is not always feasible; therefore, we need to consider alternative ways that incorporate unlabeled data as well. In recent years, Deep reinforcement learning (DRL) has gained great success in several application domains. It is an applicable method for IoT and smart city scenarios where auto-generated data can be partially labeled by users' feedback for training purposes. 

In this paper, we propose a semi-supervised deep reinforcement learning model that fits smart city applications as it consumes both labeled and unlabeled data to improve the performance and accuracy of the learning agent. The model utilizes Variational Autoencoders (VAE) as the inference engine for generalizing optimal policies. To the best of our knowledge, the proposed model is the first investigation that extends deep reinforcement learning to the semi-supervised paradigm. As a case study of smart city applications, we focus on smart buildings and apply the proposed model to the problem of indoor localization based on BLE signal strength. Indoor localization is the main component of smart city services since people spend significant time in indoor environments. Our model learns the best action policies that lead to a close estimation of the target locations with an improvement of 23\% in terms of distance to the target and at least 67\% more received rewards compared to the supervised DRL model.

\end{abstract}

}

\maketitle

\IEEEdisplaynotcompsoctitleabstractindextext
\IEEEpeerreviewmaketitle

\section{Introduction}\label{sec:Introduction}

The rapid development of Internet of Things (IoT) technologies motivated researchers and developers to think about new kinds of smart services that extract knowledge from IoT generated data. The scarcity of labeled data is a main issue for developing such solutions especially for IoT applications where a large number of sensors participate in generating data without being able to obtain class labels corresponding to the collected data. Smart cities as a prominent application area of the IoT should provide a range of high-quality smart services to meet the citizen's needs \cite{ala2015internet}. Smart buildings are one of the main building blocks of smart cities as citizens spend a significant part of their time indoors. People nowadays spend over 87\% of their daily lives indoors \cite{klepeis2001national} for work, shopping, education, etc. Therefore, having a smart environment that provides services to meet the needs to its inhabitants is a valuable asset for organizations. Such services facilitate the development of smart cities. Location-aware services in indoor environments play a significant role in this era. Examples of applications of such services are smart home management~\cite{mainetti2015location}, delivering cultural contents in museums \cite{alletto2016indoor}, location-based authentication and access control \cite{moreno2014new}, location-aware marketing and advertisement \cite{sunkada2012system}\cite{dickinson2016indoor}, and wayfinding and navigation in smart campuses \cite{torres2015enhancing}. Moreover, locating users in indoor environments is very important for smart buildings because it serves as the link that enables the users to interact with other IoT services \cite{zafari2016microlocation}. 

Deep learning is a powerful machine learning approach that provides function approximation, classification, and prediction capabilities. Reinforcement learning is another class of machine learning approaches for optimal control and decision making processes where a software agent learns  an optimal policy of actions over the set of states in an environment. In the applications where the number of states is very large, a deep learning model can be used to approximate the action values (i.e., how good an action is in a given state).
Systems that combine deep and reinforcement learning are in their initial phases but already produced competitive results in some application areas (e.g., video games). Moreover, learning approaches with no or little supervision are expected to get more momentum in the future \cite{lecun2015deep} mimicking the natural learning processes of humans and animals.

IoT applications can benefit from the decision process for learning purposes. For example, in the case of location-aware services, location estimation can be seen as a decision process in which a software agent determines the exact or closest point to a specific target. In this regard, reinforcement learning~\cite{sutton1998reinforcement} can be exploited to formulate and solve the problem. In a reinforcement learning solution, a software agent interacts with the environment and changes the state of the environment by performing some actions. Depending on the performed action, the environment sends a reward to the agent. The agent tries to maximize its rewards over time by choosing those actions that result in higher rewards. A new variation of reinforcement learning, deep reinforcement learning recently was demonstrated by Google to achieve high accuracy in the Atari games \cite{mnih2013playing} and is a suitable candidate for the learning process in the IoT applications.

In this research, we propose a semi-supervised deep reinforcement learning model to benefit from the large number of unlabeled data that are generated in IoT applications. 

The points that motivate us for this study are:
\begin{itemize}

\item In the world of IoT where sensors generate a lot of data that cannot be labeled manually for training purposes, semi-supervised approaches are valuable approaches. Moreover, building a deep reinforcement learning framework that works in a semi-supervised manner can serve many IoT applications. 

\item Games demonstrated significant improvements using DRL \cite{mnih2013playing}. 
IoT applications also can be seen as a game where the goal is to estimate the correct classification of a given input and hence can benefit from a DRL approach.

\item The learning process for the scale of smart cities requires many efforts including data gathering, analysis and classification. The strength of deep learning models stems from the latest advancements in computational and data storage capabilities. Such models can be utilized to develop scalable and efficient learning solutions for smart cities from crowd-sensed big data.

\item Smart city applications can be trained in a lab and deployed in a real environment without losing performance. For example, a self-driving car needs to learn how to perform in a variety of conditions (e.g., approaching pedestrians, handling traffic signs, etc.) which can be learned in a few test drives. But it is impossible to account for all the scenarios that might happen in a given city. 
\end{itemize}

Also, our study is motivated by specific observations regarding the localization problem including:
\begin{itemize}
\item While WiFi fingerprinting has been studied widely in the past decade for indoor positioning and the accuracy is in the range of 10 m, BLE is in its infancy for indoor localization and has yielded more fine-grained results~\cite{faragher2015location}.  	

\item There are many practical applications that need an efficient mechanism for positioning in small scale environments such as robotic soccer games to locate the position of the ball, or navigating robots in a building. Our proposed approach can be extended and used in such scenarios aiming to enhance micro-localization accuracy in support of smart environments \cite{zafari2016microlocation}. 

\end{itemize}

The contributions of this paper are as follows: 
\begin{itemize}

\item We propose a semi-supervised deep reinforcement learning framework based on deep generative models and reinforcement learning that combines the strengths of deep neural networks and statistical modeling of data density in a reinforcement learning paradigm. To the best of our knowledge, this work is the first attempt to address semi-supervised learning through deep reinforcement learning.  

\item We leverage both labeled and unlabeled data in our model. Since unlabeled data are more prevalent, this is a key feature for IoT applications where IoT sensors generate large volumes of data while they cannot be labeled easily.  Therefore, our approach helps to alleviate having a lot of labeled data. In addition, the performance of deep reinforcement learning is enhanced by using the proposed semi-supervised approach.

\end{itemize}

The idea of extending reinforcement learning algorithms to semi-supervised reinforcement learning has not been studied well so far. There are some suggestions that explain the possibility of semi-supervised reinforcement learning by having unlabeled episodes in which the agent does not receive its rewards from the environment \cite{christiano2016} \cite{amodei2016concrete}. But there is no implementation of such extensions so far. Our proposed semi-supervised deep reinforcement learning, however, follows a different approach where we incorporate a variational autoencoder \cite{kingma2014semi} in our framework as the semi-supervised module to infer the classification of unlabeled data and incorporate this information along with the labeled data to optimize its discriminating boundaries.

To apply the proposed model on a smart city scenario, we chose to perform the experiments on smart buildings which play a significant role in smart cities. Our experimental results assert the efficiency of the proposed semi-supervised DRL model compared to the supervised DRL model. Specifically, the results have been improved by 23\% for a small number of training epochs. Also, considering the average performance of both models in terms of received rewards, the semi-supervised model outperforms the supervised model by obtaining twice as many rewards.

The rest of this paper is organized as follows. Section \ref{sec:relatedWorks} organizes the recent related works into two parts: one that reviews attempts that utilize DRL models and the other that deals with the indoor localization as a case study. Section \ref{sec:background} presents related background and then introduces the details of the proposed approach. Section \ref{sec:usecase} presents a use-case study in which the proposed model is used for indoor localization systems using iBeacon signals. Experimental results are presented in Section \ref{sec:experimentalResults} followed by concluding remarks in Section \ref{sec:Conclusion}.

\section{Related Work}\label{sec:relatedWorks}

In the following sub-sections, we first review some recent research efforts that utilize deep reinforcement learning. Then, we address the latest research efforts that address the indoor localization problem from machine learning perspective. 

\subsection{Deep Reinforcement Learning} \label{ssec:related_work_DRL}

Deep reinforcement learning has been proposed in recent years \cite{mnih2013playing} and is gaining attention to be applied in various application domains.  
In the following paragraphs, we review some of the latest research efforts that utilize DRL in different application areas.

Nemati \textit{et al.} \cite{nemati2016optimal} utilized a deep reinforcement learning algorithm to learn actionable policies for administering an optimal dose of medicines like heparin for individuals. They used a sample dataset of dosage trials and their outcomes from a bunch of electronic medical records. In their model, they used a discriminative Hidden Markov Model (HMM) for state estimation and a Q-network with two layers of neurons. The medication dosage agent tries to learn the optimal policy by maximizing its total reward which is the overall fraction of time when patients are in their therapeutic activated Partial Thromboplastin Time (aPTT) range.

Deep reinforcement learning has been also applied for vehicle image classification as reported in \cite{zhao2016deepRL}. In that work, the authors propose a Convolutional Neural Network (CNN) model combined with a reinforcement learning module to guide where to look in the image for the key parts of a car. The information entropy of the classification probability of a focused image that is produced by their CNN model is considered as the reward for the reinforcement learning agent to learn to identify the next visual attention area in the image. The work in \cite{caicedo2015active} also reports on object localization in images by focusing attention on candidate regions using a deep reinforcement learning approach. In \cite{zhu2016target}, a visual navigation application is presented that uses a variation of deep reinforcement learning by which robots can navigate in a space toward a visual target. 

Li \textit{et al.} \cite{li2016traffic} also developed a deep reinforcement learning approach for traffic signal timing aiming to have a better signal timing plan. In their model which consists of a four-layer stacked autoencoder neural network to estimate the Q-function, they use the queuing lengths of eight incoming lanes to an intersection as the state of the system at each time. They also define two actions: stay in the current traffic lane, or change the lane to allow other traffic to go through the intersection. The absolute value of the difference between the length of opposite lanes serves as the reward function. Their results show that their proposed model outperformed other conventional reinforcement learning approaches. 

Resource management is another task that can use DRL as its underlying mechanism. In their report, Mao \textit{et al.} \cite{mao2016resource} formulate the problem of job scheduling with multiple resource demands as a deep reinforcement learning process. In their approach, the objective is to minimize the average job slowdown. The reward function was defined based on the reciprocal duration of the job in order to guide the agent toward the objective.

Another application in which deep reinforcement learning played a key role is natural language understanding for text-based games \cite{narasimhan2015language}\cite{he2015deep}. For example, Narasimhan \textit{et al.} \cite{narasimhan2015language} used Long Short-Term Memory (LSTM) networks to train the agent with useful representations of text descriptions and a Deep Q-Network (DQN) to approximate Q-functions. Other disciplines like energy management also have incorporated DRL to improve energy utilization \cite{vinf2016deer}. 

\subsection{Review of Indoor Localization}\label{ssec:related_work_usecase}

To the best of our knowledge, there are no prior research efforts that utilized DRL for localization. In the following paragraphs, we review the different machine learning approaches that were utilized in the recent research literature to provide indoor localization services.

Among the approaches for deploying location-based services, Relative Signal Strength (RSS) fingerprinting is one of the most promising approaches. However, there are some challenges that need to be considered in the deployment of such approach including fingerprint annotation and device diversity \cite{wang2016indoor}. The use of fingerprint-based approaches to identify an indoor position has been studied well in the past decade. Researchers have studied different machine learning approaches in this context including: SVM, KNN, Bayesian-based filtering, transfer learning, and neural networks \cite{zhang2016deep}.

It has been shown that for coarse-grained positioning applications based on Bluetooth Low Energy (BLE) RSS fingerprinting, the estimation to decide if a device is inside a room or not yields pretty reliable results \cite{kajioka2014experiment}. 

The authors in \cite{wang2013bluetooth} report their experimental indoor positioning results based on BLE RSS readings. They study the accuracy of three methods including Least Square Estimation (LSE), Three-border positioning, and Centroid positioning. In their testing area, which is a $6\times 8$ square meters classroom with four BLE stations at the corners, the LSE algorithm shows more accurate positions compared to the two other methods. However, the overall accuracy of these three algorithms is satisfactory.

Museums are good environments for using BLE to provide location-awareness since usually the building and its contents do not allow changes due to preservation policies. The authors in \cite{alletto2016indoor} developed a system to make interactive cultural displays in a museum with BLE beacons combined with an image recognition wearable device. The wearable device performs localization by receiving BLE signals from the beacons to identify the room in which it is located. It also identifies artworks by an image processing service. The combination of the closest beacon identifier and the artwork identifier are fed to a processing center to retrieve the appropriate cultural content.

In \cite{wang2015deepfi}, the authors present a system called DeepFi that utilizes a deep learning method over fingerprinting data to locate indoor positions based on channel state information (CSI). As many other fingerprinting approaches, their system consists of offline training and online localization phases. In the off-line training phase, they exploit deep learning to train all the weights as fingerprints based on the previously stored CSI. Their evaluations in a living room and laboratory settings show that the use of deep learning result in improved localization accuracy of 20\%. While their use of CSI approach is limited to WiFi networks, not all available Network Interface Cards (NICs) in the market support obtaining measurements from the different network channels. 

In \cite{gu2015semi}, deep learning joined with semi-supervised learning as well as extreme learning machine are applied to unlabeled data to study the performance of feature extraction and classification phases of indoor localization. In their study, deep learning network and semi-supervised learning generate high level abstract features and more accurate classification while extreme learning machine can speed up the learning process. Their test setting is a 10$\times$15 square meters area. Their results show that deep learning can improve the accuracy of fingerprinting by at least 1.3\% for the same training dataset compared to a shallow learning method. Also increasing unlabeled data has a positive effect on the accuracy compared to shallow feature methods. Compared to other deep learning methods including stacked autoencoder, deep belief network, and multi-layer extreme learning machine, their approach improves the accuracy at least 10\%. 

In another study \cite{zhang2016deep}, the authors propose a WiFi localization approach using deep neural networks (DNN). In their system, a four-layer deep learning model is used to extract features from WiFi RSS data. In their approach, the authors use Stacked Denoising Autoencoder and Backpropagation for the training steps. In the online positioning phase, the estimated position based on DNN is refined by an HMM component. Their experiments assert that the number of hidden layers and neurons have a direct effect on the localization accuracy. Increasing the layers leads to better results, but at some point when the network is made deeper, the results start degrading. Their result shows that when using three hidden layers with 200 neurons for each layer, the model achieves the best accuracy.

Ding \textit{et al.} \cite{ding2013fingerprinting} also used an Artificial neural network (ANN) for WiFi fingerprinting localization. They proposed a localization approach that uses ANNs in conjunction with a clustering method based on affinity propagation. By affinity propagation clustering, the training of the ANN model has been faster and the memory overhead has been lowered. They also reported improved positioning accuracy compared to other baseline methods.

In \cite{luo2016deep}, a deep belief network (DBN) is used for a localization approach that is based on fingerprinting of ultra-wideband signals in an indoor environment. Parameters of channel impulse response are used to get a dataset of fingerprints. Compared to other methods, the author demonstrated that DBN can improve the localization accuracy.

The work in \cite{zhang2016device} also reports using a deep learning model in conjunction with a regression model to automatically learn discriminative features from the received wireless signal. The authors also use a softmax regression algorithm to perform device-free localization and activity recognition. They report that their proposed method can improve the localization accuracy by 10\% compared to other methods.

Semi-supervised algorithms have also been widely applied to the localization problem to utilize the unlabeled data for the prediction of an unknown location. For example, in \cite{pulkkinen2011semi} the authors presented a semi-supervised algorithm based on the manifold assumption to obtain tagged fingerprints out of unlabeled data using a small amount of labeled data.  They map the high-dimensional space of fingerprints into a two-dimensional space and achieved an average error of 2 meters. 

Our work in this paper presents several significant differences compared to the aforementioned approaches. First, related research studies in deep reinforcement learning do not exploit the statistical information of unlabeled data, while our proposed DRL approach is extended to be semi-supervised and utilizes both labeled and unlabeled data. Second, these approaches provide an application-dependent solution, while our work is a general framework that can work for a variety of IoT applications. Third, for localization systems, all aforementioned deep learning solutions rely on WiFi fingerprinting, while the context of BLE fingerprinting has not been studied in conjunction with deep learning or reinforcement learning approaches.

\section{Background and Proposed Approach}\label{sec:background}

In the following  sub-sections, we first describe the fundamentals of variational autoencoders. Then we describe our proposed semi-supervised DRL model by adopting a variational autoencoder in a deep reinforcement learning model. We develop the theoretical foundation of our method based on \cite{kingma2014semi} and \cite{mnih2013playing}.

\subsection{Semi-Supervised Learning Using VAE}

Semi-supervised learning methods aim to improve the generalization of supervised learning tasks using unlabeled data \cite{weston2012deep}. They usually use a small set of annotated data along with a larger number of unlabeled data to train the model.
In a semi-supervised setting, we have two datasets; one is labeled and the other is unlabeled. The labeled dataset is denoted by $X_l =  (x_1,\ldots, x_l)$ for which labels $Y_l = (y_1, \ldots, y_l)$ are provided. The other set is $X_u = (x_{l+1},\ldots, x_{l+u})$ with unknown labels.
Semi-supervised algorithms are built based on at least one of the following three assumptions \cite{chapelle2006semi}: The smoothness assumption, states that if two points $x_1$ and $x_2$ are close to each other, then their corresponding labels are very likely to be close to each other. The cluster assumption implies how to identify discrete clusters. It states that if two points are in the same cluster, it is more probable that they have the same class label. The manifold assumption points out that high dimensional data can be mapped to a lower dimensional one (i.e., the principle of parsimony) such that the supervised algorithm still approximates the true class of a data point.

For the semi-supervised part of our proposed model, we adopt the deep generative model based on variational autoencoders (VAE) \cite{kingma2014semi}. This model has been used for semi-supervised tasks such as the recognition of handwritten digits, house number classification and motion prediction \cite{walker2016uncertain} with impressive results. Figure \ref{fig:vae} shows the structure of a typical VAE model.
For each data point $x_i$ there is a vector of corresponding latent variables denoted by $z_i$.
The distribution of labeled data is represented by $\tilde{p_l}(x, y)$, while unlabeled data are represented by $\tilde{p_u}(x)$.

The latent feature discriminative model (M1) is created based on:

\begin{figure}
	
	\begin{center}
		
		\includegraphics[width=0.45\textwidth]{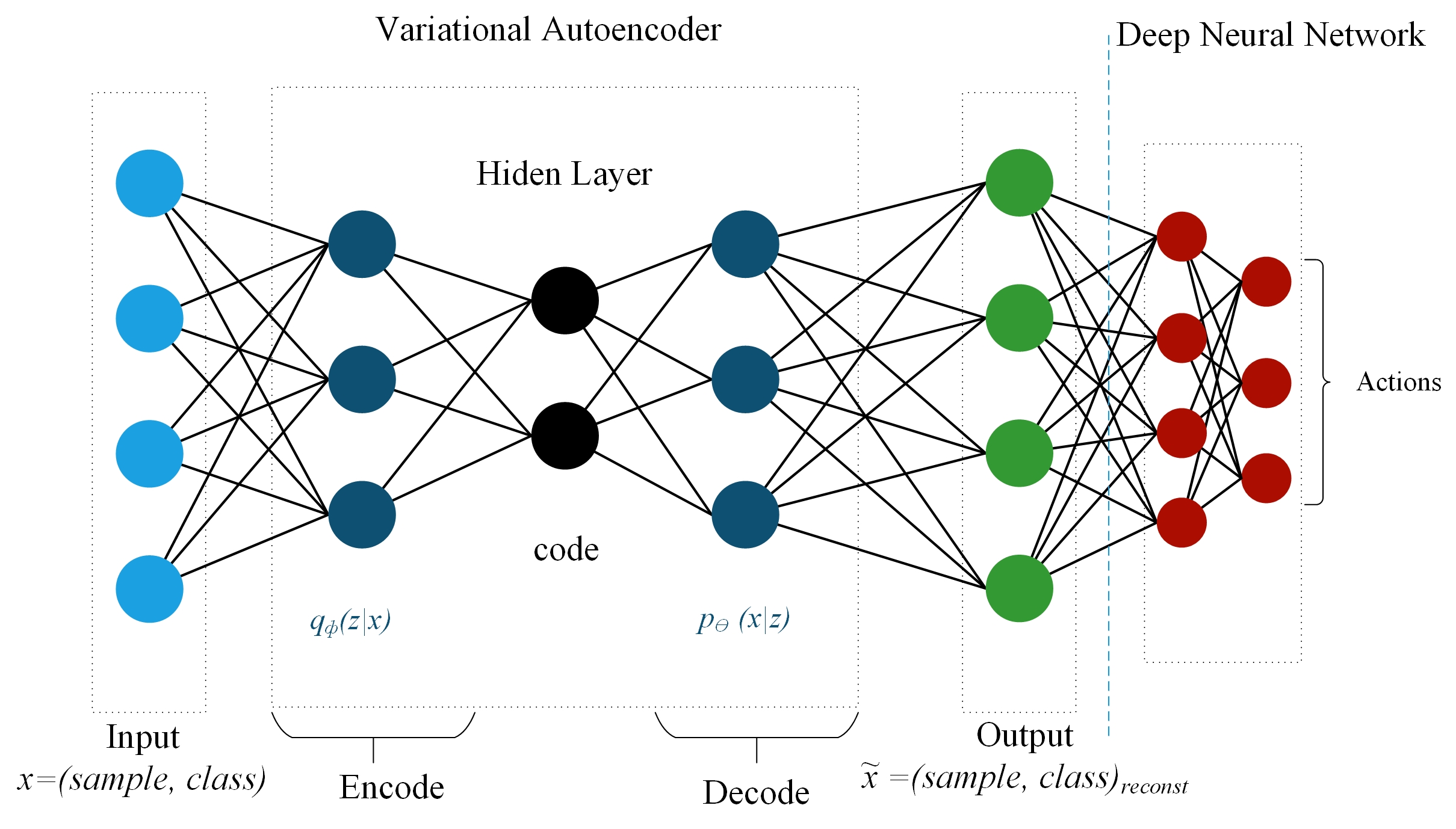}
		
	\end{center}
	
	\caption{The high-level concept of a variational autoencoder adopted for deep reinforcement learning. 
    }\label{fig:vae}
	
\end{figure}

\begin{equation}
	p(z) = \mathcal{N}(z|0,I); \  \  \  \  p_{\theta}(x|z)=f(x;z,\theta),
\end{equation}
in which $p(z)$ is Gaussian distributed with mean vector $0$ and variances presented in an identity matrix $I$. The function $f(x;z,\theta)$ is a nonlinear likelihood function with parameter $\theta$ for latent variable $z$ based on a deep neural network. 

The generative semi-supervised model for generating data using a latent class variable $y$, in addition to a latent variable $z$ is (M2):

\begin{equation}
\begin{gathered}
p(y) = Cat(y|\pi); \ \ p(z) = \mathcal{N}(z|0,I); \\  p_{\theta}(x|y,z)=f(x;y,z,\theta),
\end{gathered}
\end{equation}
where $Cat(y|\pi)$ represents a categorical distribution or in general a multinomial distribution with a vector of probabilities $\pi$ whose elements sum up to 1. In the dataset, if no label is available, the unknown labels $y$ are considered as latent variables in addition to $z$.

The models have two lower bound objectives. To describe the model objectives, a fixed form distribution $q_{\phi}(z|x)$ is introduced with parameter $\phi$ that helps us to estimate the posterior distribution $p(z|x)$.
For all latent variables in the models, an inference deep neural network is introduced to generate a distribution of the form $q_{\phi}(.)$. For M1, a Gaussian inference network $q_{\phi}(z|x)$  is used for latent variable $z$:   

\begin{equation}
\begin{gathered}
q_{\phi}(z|x) = \mathcal{N}(z|\mu_{\phi}(x),diag(\sigma^2_{\phi}(x))),
\end{gathered}
\end{equation}
in which $\mu_{\phi}(x)$ is the vector of means, $\sigma_{\phi}(x)$ is the vector of  standard deviations, and $diag$ creates a diagonal matrix. 
For M2, an inference network is used for latent variables $z$ and $y$ using Gaussian and multinomial distributions, respectively: 

\begin{equation} \label{eq:M2inference}
\begin{gathered}
q_{\phi}(z|y,x) = \mathcal{N}(z|\mu_{\phi}(y,x),diag(\sigma^2_{\phi}(x)));\\ q_{\phi}(y|x) = Cat(y|\pi_{\phi}(x)),
\end{gathered}
\end{equation}
where $\pi_{\phi}(x)$ is a vector of probabilities.

The lower bound for M1 is: 

\begin{equation}
\begin{gathered}
\log p_{\theta}(x) \geq  \mathbb{E}_{q_{\phi}(z|x)}
[\log p_{\theta}(x|z)] - KL[q_{\phi}(z|x) || p_{\theta}(z)]\\ = -\mathcal{J}(x),
\end{gathered}
\end{equation}  
in which $KL$ is the Kullback-Leibler divergence function between the encoding and prior distribution and can be obtained as $KL(q_{\phi}\|p_{\theta})=\sum_{i} q_{\phi_i} \log\dfrac{q_{\phi_i}}{p_{\theta_i}}$.

For the model M2, two cases should be considered. The first one deals with labeled data:

\begin{equation}
\begin{gathered}
\log p_{\theta}(x, y) \geq \mathbb{E}_{q_{\phi}(z|x, y)}
[\log p_{\theta}(x|y, z)] + \log p_{\theta}(y) \\+ \log p_{\theta}(z) - \log q_{\phi}(z|x,y)] = -\mathcal{L}(x, y).
\end{gathered}
\end{equation}  

When dealing with unlabeled data, $y$ is treated as a latent variable and the resulting lower bound is:
\begin{equation}
\begin{gathered}
\log p_{\theta}(x) \geq \mathbb{E}_{q_{\phi}(y, z|x)}
[\log p_{\theta}(x|y, z)] + \log p_{\theta}(y) \\ + \log p_{\theta}(z) - \log q_{\phi}(y, z|x)] 
= -\mathcal{U}(x).
\end{gathered}
\end{equation}
Then the whole dataset has its bound of marginal likelihood as:
\begin{equation}
\mathcal{J} = \sum_{(x,y) \sim \tilde{p_{l}}} \mathcal{L}(x,y) + \sum_{x \sim \tilde{p_{u}}} \mathcal{U}(x)
\end{equation}

By adding a classification loss to the above function, the optimized objective function becomes:

\begin{equation}
\mathcal{J^{\alpha}} = \mathcal{J} + \alpha \ . \ \mathbb{E}_{\tilde{p_l}(x,y)} [-\log q_{\phi}(y|x)],
\end{equation}
where $\alpha$ adjusts the contributions of the generative and discriminative models in the learning process. During the training process for both models M1 and M2, the stochastic gradient of $\mathcal{J}$ is computed at each minibatch to be used for updating the generative parameters $\theta$ and the variational parameters $\phi$. 

\subsection{Semi-Supervised Deep Reinforcement Learning}\label{sec:problemForm}

To adopt a deep reinforcement learning approach, we need to define the following elements for a Markov Decision Process (MDP). The goal of the MDP in a reinforcement learning problem is to maximize the earned rewards.

\textit{Environment}: The environment is the territory that the learning agent interacts with.

\textit{Agent}: The agent observes the environment, receives sensory data and performs a valid action. It then receives a reward for its action. Through training, the agent learns to maximize its rewards.

\textit{States}: The finite set of states that the environment can assume. Each action of the agent puts the environment in a new state. 

\textit{Actions}:  The finite set of available actions that the agent can perform causing a transition from state $s_t$ at time $t$ to state $s_{t+1}$ at time $t+1$.

\textit{Reward function}: This function is the immediate feedback for performing an action. The reward function can be defined such that it reflects the closeness of the current state to the true class label; i.e., $r(s_t,a_t,s_{t+1},y) = closeness(s_{t+1},y)$. Depending on the problem, different distance measurements can be applied. The point is that we need to devise larger positive rewards for more compelling results and negative rewards for distracting ones. 

\textit{State transition distribution}: is the probability that action $a$ in state $s$ at time $t$ will lead to state $s'$ at time $t+1$: $P_a(s, s') = Pr (s'|s,a)$. 

Having these components, the main problem is to find a policy $\pi$ (where $\pi = a_t$) that maximizes the rewards: 
$R_t = \sum_{t=0}\gamma^t r_{t}$, 
in which $\gamma$ is a discount factor $0\leq\gamma<1$.

In the deep Q-Network approach, we need a deep neural network that approximates the optimal action-value function (Q) \cite{mnih2015human}:

\begin{equation}
	Q^*(s,a) = \max_{\pi} \mathbb{E}[r_t+\gamma r_{t+1}+\gamma^2 r_{t+2} + \ldots | s_t = s, a_t = a, \pi]
\end{equation}

This function finds the maximum sum of rewards $r_t$ discounted by $\gamma$ at each time-step $t$, achievable by a behavior policy $\pi = P(a|s)$, after making an observation ($s$) and taking an action ($a$). We can convert this equation to a simpler approximation function using \textit{Bellman equation}. 
For a sequence of states $s'$ and for all possible actions $a'$, if the optimal value $Q^*(s', a')$ is known, then we can obtain the optimal strategy by selecting the action $a'$ that maximizes the expected value of $r+\gamma Q^*(s', a')$: 

\begin{equation}
\begin{gathered}
	Q^*(s,a) = \mathbb{E}_{s'}\left[r+\gamma \max_{a'}Q^*(s',a') \vert s,a \right]
\end{gathered}
\end{equation}

To estimate the optimal action-value function, we use a non-linear function approximator (i.e., a neural network with weights $\theta$) such that $Q(s,a;\theta) \approx Q^*(s,a)$. The network can be trained by minimizing the loss functions $L_i(\theta_i)$ that is updated at each time-step.  

We perform experience replay, so we keep track of the agent's experiences $e_t = (s_t,a_t,r_t,s_{t+1})$ at each time-step $t$ in a replay dataset $D_t = \{e_1,\ldots,e_t\}$. This dataset of recently experienced transitions along with the experience replay mechanism are critical for the integration of reinforcement learning and deep neural networks \cite{mnih2015human}. 

Q-learning updates are applied on samples from the training data $(s,a,r,s')$ that are uniformly drawn from the experience replay storage $D$. The Q-learning update in iteration $i$ uses the following loss function:
\begin{equation}
\begin{gathered}
	L_i(\theta_i) = \mathbb{E}_{(s,a,r,s') \sim  U(D)}\bigl[\bigl(\overbrace{r+\gamma \max_{a'} Q(s',a';\theta_{i-1})}^{\xi_i}\\ - Q(s, a; \theta_i) \bigr)^2\bigr]
\end{gathered}
\end{equation}
in which $\theta_i$ represents the network parameters in iteration $i$, and the previous network parameters $\theta_{i-1}$ are used to compute the target ($\xi_i$).
The gradient of the loss function is computed with respect to the weights of the network:
\begin{equation}\label{eq:gradient}
\begin{gathered}
\nabla_{\theta_i} L_i(\theta_i) = \mathbb{E}\bigl[\bigl(r+\gamma \max_{a'} Q(s',a';\theta_{i-1}) - Q(s, a; \theta_i)\bigr)\\  \nabla_{\theta_i} Q(s,a; \theta_i)\bigr]
\end{gathered}
\end{equation}

\begin{figure*}
	
	\begin{center}
		
		\includegraphics[width=1\textwidth]{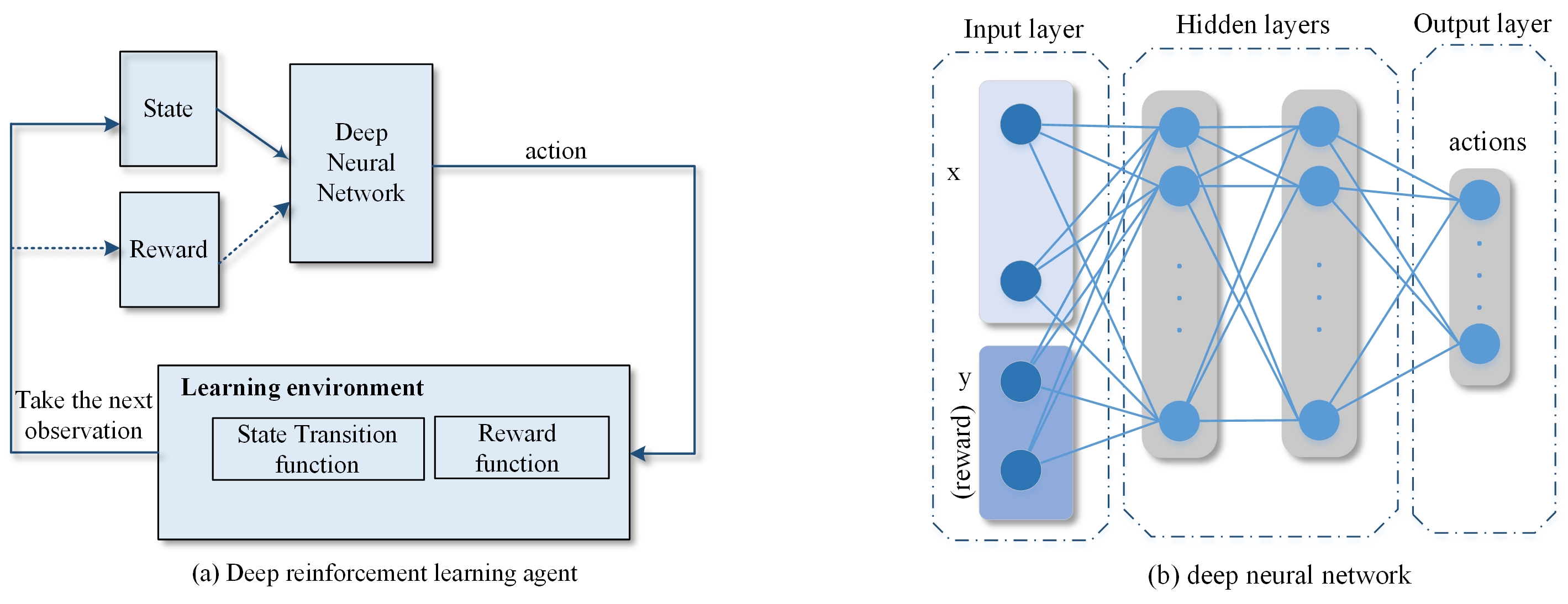}
		
	\end{center}
	
	\caption{The proposed model. (a) The DRL agent considers x values as the next state of the environment and y values as a mechanism to compute the reward. For unlabeled data, the x values are only incorporated into the model. (b) a general deep neural network to be used for supervised DRL.}\label{fig:model}
	
\end{figure*}

The semi-supervised DRL algorithm is then  described in Algorithm \ref{alg:drl_algorithm1} to learn from both labeled and unlabeled data. 

\begin{algorithm}
	
	\caption{Semi-Supervised DRL Algorithm}
	
	\label{alg:drl_algorithm1}
	
	\begin{algorithmic}[1]
		
		\STATE{Input: A dataset of labeled and unlabeled data $\{(X_l, Y_l), X_u\}$}

		\STATE{Initialize the model parameters $\theta$, $\phi$, environment, state space, and replay memory $\mathcal{D}$}
		\FOR{$episode \gets 1 \textrm{ to } M$}
			\FOR{\textbf{each} sample  $(x,y)$ or $x$ in dataset}
				\STATE{ $s_0 \gets$ make observation of sample $x$}
                \FOR{$t \gets 0 \textrm{ to } T$}
					\STATE{ Take an action $a_t$ using $\epsilon$-greedy strategy}
					\STATE{ Perform action $a_t$ to change the current state $s_t$ to the next state $s_{t+1}$}
                    \IF{sample is unlabeled}
                    \STATE{Infer the label based on (\ref{eq:M2inference}): $q_{\phi}(y|x)$ and get approximate reward $r_t=closeness(s_{t+1}, y)$}
                    \ELSE 
                    \STATE{Observe reward $r_t$ that corresponds to label $y$}
                    \ENDIF
					\STATE{ Store transition ($s_t$, $a_t$, $r_t$, $s_{t+1}$) in $\mathcal{D}$ }
                    \STATE{ Take a random minibatch of transitions ($s_k$, $a_k$, $r_k$, $s_{k+1}$) from $\mathcal{D}$; $0 < k\leq length(minibatch)$  }
                    \IF {$s_{k+1}$ is a terminal state}
                    \STATE {$\xi_k$ = $r_k$}
                    \ELSE \STATE {$\xi_k$ = $r_k + \gamma \max_{a'} Q(s_{k+1},a';\theta)$}
                    \ENDIF
                    \STATE {Apply gradient descent on $(\xi_k - Q(s_k, a; \theta))^2$ based on (\ref{eq:gradient})}
                    \ENDFOR
                    \STATE{\textbf{end for}}
			\ENDFOR
            \STATE{\textbf{end for each} }
		\ENDFOR	
        \STATE{\textbf{end for} }
		
	\end{algorithmic}
	
\end{algorithm}

Figure \ref{fig:model} shows the high-level model that uses the deep reinforcement learning technique in conjunction with a generative semi-supervised model instead of a DNN (c.f.  Figure~\ref{fig:model}-b) to handle unlabeled observations. The VAE is extended to have an additional hidden layer and an output to generate the actions.

As other learning processes, the training process for this algorithm is performed offline while policy prediction is performed online. Hence, the algorithm can handle problems with high-dimensional and high-volume data using high performance computing facilities (e.g., cloud servers) to generate the model for online policy prediction. This ability stems from the integration of deep neural networks with reinforcement learning to generate approximation functions for high-dimensional datasets. The performance of this integrated model outperforms the traditional methods of reinforcement learning.

\section{Use Case: Indoor Localization}\label{sec:usecase}

Several use cases can be envisaged of the proposed approach in a smart city context. For example, this approach can be used for home energy management in conjunction with the Non-Intrusive Load Monitoring (NILM) method and smart meters. In such systems, a small set of labeled data provides individual appliances' usages and their on and off times. A semi-supervised deep reinforcement learning model can be trained over this small-scale training dataset as well as the stream of unlabeled data with the objective of optimizing energy usage by controlling when to switch appliances on and off.

It can also be used in the context of Intelligent Transportation Systems (ITS) by smart vehicles for navigation in a city context. In such applications, a combination of several factors can be used for the reward function such as closeness to the destination, shortest path, speed, speed variability, etc. The vehicle needs to be trained on several test drives then it uses the large set of unlabeled data to accurately navigate through the city.

Due to the importance of indoor localization and ease of implementation, we showcase the proposed method on the localization problem in the context of smart campus, which is part of a larger smart city context. Despite the fact that indoor localization has been studied extensively in recent years, still it is an open problem bringing several challenges that need to be tackled. 

Indoor positioning systems have been proposed with different technologies such as vision, visual light communications (VLC), infrared, ultrasound, WiFi, RFID, and BLE \cite{mainetti2014survey}. One determining factor for organizations to choose a technology is the cost of the underlying technologies and devices. Among the aforementioned technologies, BLE is a low-cost solution that has attracted the attention for academic and commercial applications \cite{zafari2016microlocation}. A combination of BLE and iBeacon technologies to design an indoor location-aware system brings many advantages to buildings that are not equipped with Wireless networks. Since iBeacons devices are of a small form factor, they can be deployed quickly and easily without changing or even tapping into the building's electrical and communications infrastructure \cite{mainetti2014survey}. 

In recent years, deep learning has been shown to perform favorably compared to other machine learning approaches. One main challenge for deep learning is the need to collect a large volume of labeled data (a.k.a calibration procedure). Typically, scanning a large-scale area like a city or a campus to collect unlabeled data is fairly straightforward. Therefore, to benefit from the enormous volume of unlabeled data, we apply the semi-supervised deep reinforcement learning approach to investigate the benefits of unlabeled data in practical scenarios.

Compared to many related works that have performed their studies in a simulated environment, a small area, or in an isolated testbed, we conducted our experiments in an academic library that is a large and busy operational environment where thousands of visitors commute every day. So it is a valuable experiment that can be beneficial for the IoT and AI communities. In addition, there are no similar attempts that address the positioning problem through the reinforcement learning approach.

In this case study, we utilize a grid of iBeacons to implement a location-aware service offering in a campus setting. In our work, we use the iBeacons' Received Signal Strength Indicator (RSSI) as the raw source of input data for a deep reinforcement learning model to identify indoor locations.

RSSI is usually represented by a negative number between 0 and -100 and in localization systems it can be used as an indication of the distance separating the transmitter from the receiver (i.e., ranging). In addition to the separating distance, RSSI is affected by some other factors such as movement of people and objects amidst the signals, temperature and humidity of the environment. The distance estimation from a given point to an iBeacon can be derived as follows:

\begin{equation}
RSSI = -(10 × n)log_{10}(d) + A, 
\end{equation}
where $n$ is the signal propagation constant, $d$ is the distance in meters and $A$ is the offset RSSI reading at 1 meter from the transmitter. 

Due to fluctuations of the received signal strength, many research studies that utilize RSSI fingerprinting perform a preprocessing step to extract more representative features. Some of these preprocessing approaches include averaging multiple RSSI values for the same location, use Gaussian distribution model to filter outliers, and using PCA to reduce the effect of noise in addition to offering new features. In our work, we performed a categorization preprocessing in which a RSSI category represents a range of RSSI values. We explain the exact procedure in section \ref{ssec:byDeepLearn}.

\subsection{Description of the Environment}
The environment is represented as a set of positions that are labeled by row and column numbers. Each position is also associated with the set of RSSI values from the set of deployed iBeacons. The agent  observes the environment by receiving RSSI values at each time. Our design requires the agent to take action based on the three most recent RSSI observations.

The agent can choose one of the allowed eight actions to move in different directions. In turn, the agent obtains a positive or negative reward according to its proximity to the right point. The goal of the agent is to approximate the position of the device that has received the RSSI values from the environment by moving in different directions. 

\begin{table}
\renewcommand{\arraystretch}{1.2}
	\centering
	\caption{List of actions to perform positioning}
	\label{tbl:actions}
	\begin{tabular}{|c|c|c|c|c|c|c|c|c|}
		\hline
		\rowcolor[HTML]{A0C0C0} 
		\textbf{Action\#} & \textbf{0} & \textbf{1} & \textbf{2} & \textbf{3} & \textbf{4} & \textbf{5} & \textbf{6} & \textbf{7} \\ \hline
		\textbf{Move to}  & West       & East       & North      & South      & NW         & NE         & SW         & SE         \\ \hline
	\end{tabular}
\end{table}

To adopt a deep reinforcement learning approach, we need to define the following elements for the MDP. 

\begin{figure}
	
	\begin{center}
		
		\includegraphics[width=0.45\textwidth]{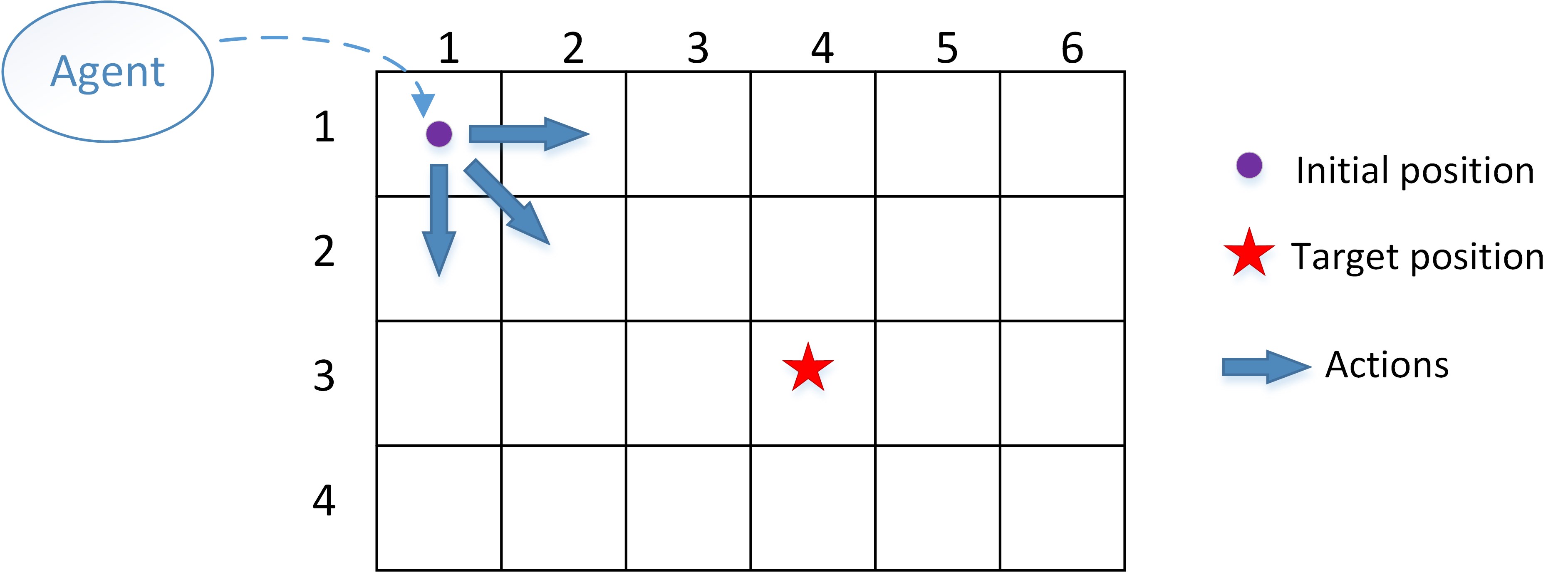}
		
	\end{center}
	
	\caption{Illustration of a typical indoor environment for deep reinforcement. learning}\label{fig:environment}
	
\end{figure}

\textit{Environment}: the active environment is a floor on which a particular position should be identified based on a vector of iBeacon RSSI values. The environment is divided into a grid of same-size cells as shown in Figure \ref{fig:environment}.

\textit{Agent}: The positioning algorithm itself is represented as an agent. The agent interacts with the environment over time. 

\textit{States}: The state of the agent is represented as a tuple of these observations:
\begin{enumerate}
\item  a vector of RSSI values,
\item current location (identified by row and column numbers), and
\item distance to the target (for labeled data).
\end{enumerate}

\textit{Actions}:  The action is to move to one of the neighboring cells in a direction of North, East, West, South or in between directions like North West (NW). The first action chooses a random state in the grid. Table \ref{tbl:actions} shows the list of allowed actions.

\textit{Reward function}: the reward function is the reciprocal of the distance error. The reward function has a positive value if the distance to the target point is less than a threshold ($\delta$). Otherwise, the agent receives a negative reward. Whenever the agent is close to the target, it gains more rewards. On the other hand, if the agent wanders away from the target and its distance is larger than a threshold ($\delta$), it gains a negative reward. The reward function is represented as follows: 

\begin{displaymath}
r_{t}  =\left\{ \begin{array}{ll}

\frac{1}{\|O_t - S_t\|} &\mbox{if 0 $<$ $\|O_t - S_t\| \leq \delta$}\\
-\|O_t - S_t\| &\mbox{$otherwise$.}

\end{array}\right.
\end{displaymath}
in which $O_t$ is the observed location and $S_t$ is the target location.

\section{Experimental Results}\label{sec:experimentalResults}
\begin{figure}
	
	\begin{center}
		
		\includegraphics[width=0.45\textwidth]{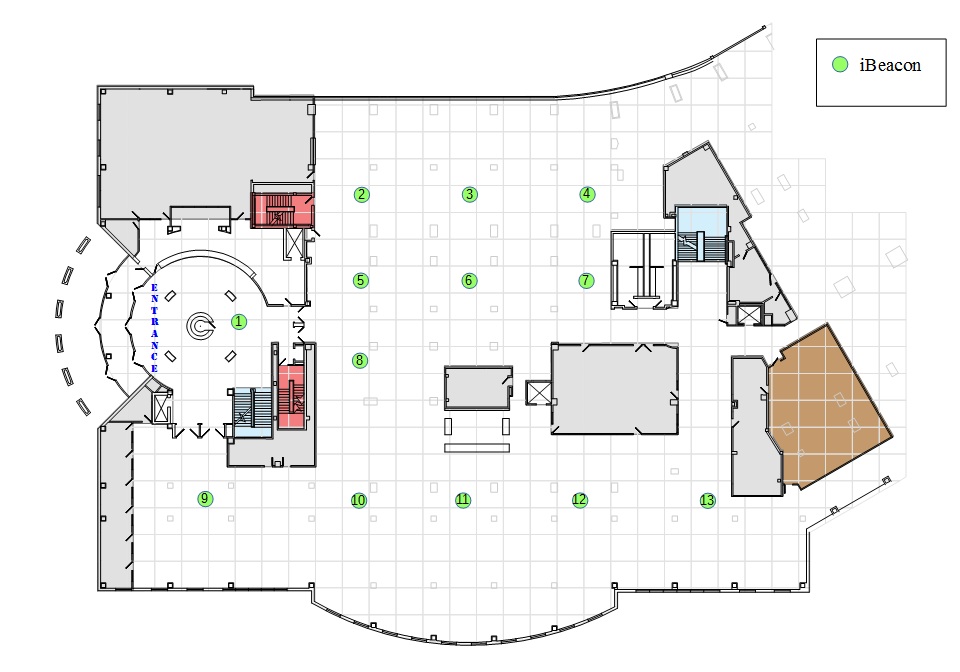}
		
	\end{center}
	
	\caption{Experimental setup with iBeacons.}\label{fig:floorplan}
	
\end{figure}

Here we describe our evaluation on a real world dataset. Our experiments were carried on the first floor of Western Michigan University (WMU) Waldo library. Figure \ref{fig:floorplan} shows the overall layout of the deployment site. In our work, we use the iBeacon RSSI values to serve as the raw source of input data to identify indoor locations.  
Smartphones are also utilized to sense the iBeacons' signals and to compute the current position of the user with respect to the set of known iBeacons. Our model utilizes the semi-supervised deep reinforcement learning algorithm to learn from the historical patterns of RSSI values and their corresponding estimated positions to improve its policy when identifying a position based on previously unseen RSSI values.

\subsection{Dataset}

Our dataset is gathered from a real-world deployment of a grid of iBeacons in a campus library area of 200 ft. $\times$180 ft. We mounted 13 iBeacons  on the ceiling of the first floor of Waldo Library at Western Michigan University which contains many pillars that might deteriorate the iBeacons signals. So we arranged the iBeacons such that we could get signal coverage by several iBeacons. Each iBeacon is separated by a distance of 30-40 ft. from adjacent iBeacons. To capture the signal strength indicator of these iBeacons, we divided the area into small zones by mapping a grid that has cells of size 10$\times$10 square ft. We also developed a specific mobile app to capture training data. For that purpose, we stood on each cell and captured all the iBeacons' received signals. We also manually assigned the location (i.e., label of the cell) to the captured signals. We stored at least three instances of RSSIs for each cell to have a more reliable measurement and consequently to reduce the effect of noisy data. Overall, we collected 820 labeled data points for training, 600 data points for testing, and 5200 data points are unlabeled for semi-supervised learning.

\subsection{
Preprocessing
}\label{ssec:byDeepLearn}

Our initial experiments with the raw RSSI values for supervised deep learning showed that the relationship between the features are not truly revealed by deep learning models. So we have enriched the features by adding two sets of features to the original features. So we have three feature sets as: 
\begin{itemize}

\item Raw: The original features that come from the direct RSSI readings. 
\item S1: The set of features that represent the mutual differences of iBeacon RSSI values; i.e., $r_i - r_j; \forall i , j \in {iBeacons}$ \& $i \neq j$, representing the difference between the RSSI value of beacon $i$ and beacon $j$. 
\item S2: The other set of features designed to represent the categorical values of RSSIs in a Boolean membership mode such that for each beacon we define several categories by a specific interval (e.g., 10) and then represent each RSSI value with the category to which it belongs.
\end{itemize}
Table \ref{tbl:featuresAccuracy} shows the average accuracy of the different feature sets during ten replications. These features are added to the raw features. As can be seen from the table, adding features set S1 to raw features has a minor effect on the average accuracy. On the other hand, adding features set S2 increases the average accuracy especially for finer grained positioning. Also, the combination of S1 and S2 is not as good as using only S2, since S1 lowers the accuracy. This observation points out that enriching a feature set by pairwise differences of RSSI values (S1) has a minor negative effect on the accuracy of the model since those features are not solid discriminative factors. 

\begin{table}
\renewcommand{\arraystretch}{1.2}
	\centering
	\caption{Accuracy of different feature sets in a deep neural network}
	\label{tbl:featuresAccuracy}
	\begin{tabular}{|c|l|c|c|c|c|}
\hline
\multirow{2}{*}{\textbf{Interval}} & \multirow{2}{*}{\textbf{feature set}} & \multicolumn{4}{c|}{\textbf{Accuracy}}                                                                  \\ \cline{3-6} 
                                   &                                       & \textbf{$\leq$ 1m}     & \textbf{$\leq$ 3m}     & \textbf{$\leq$ 6m}     & \textbf{$\leq$ 9m}     \\ \hline
-                                  & raw                                   & 0.17                      & 0.47                      & 0.74                      & 0.95                      \\ \hline
\multirow{3}{*}{10}                & raw\_s1                               & 0.18                      & 0.49                      & 0.75                      & 0.95                      \\ \cline{2-6} 
                                   & raw\_s2                               & 0.26                      & 0.55                      & 0.75                      & 0.97                      \\ \cline{2-6} 
                                   & raw\_s1\_s2                           & 0.24                      & 0.52                      & 0.74                      & 0.96                      \\ \hline
5                                  & raw\_s2                               & \multicolumn{1}{l|}{0.30} & \multicolumn{1}{l|}{0.57} & \multicolumn{1}{l|}{0.76} & \multicolumn{1}{l|}{0.97} \\ \hline
\end{tabular}
\end{table}

The table also demonstrates that using S2 features when RSSI categorical interval is set to 5 leads to even better results. 
Therefore, based on these results we use the combination of raw features and S2. Using this preprocessing, each data point $x_i$ is represented as a vector of 13 RSSI values plus 156 range membership features (i.e., 12 range for the 13 beacons) resulting in a total of 169 features: $x_i = (r_1,\ldots,r_{169})$.  Each $y_i$ is a label of ($row, col$) pointing to a specific location.

\begin{figure*}	
	\begin{center}		
		\includegraphics[width=1\textwidth]{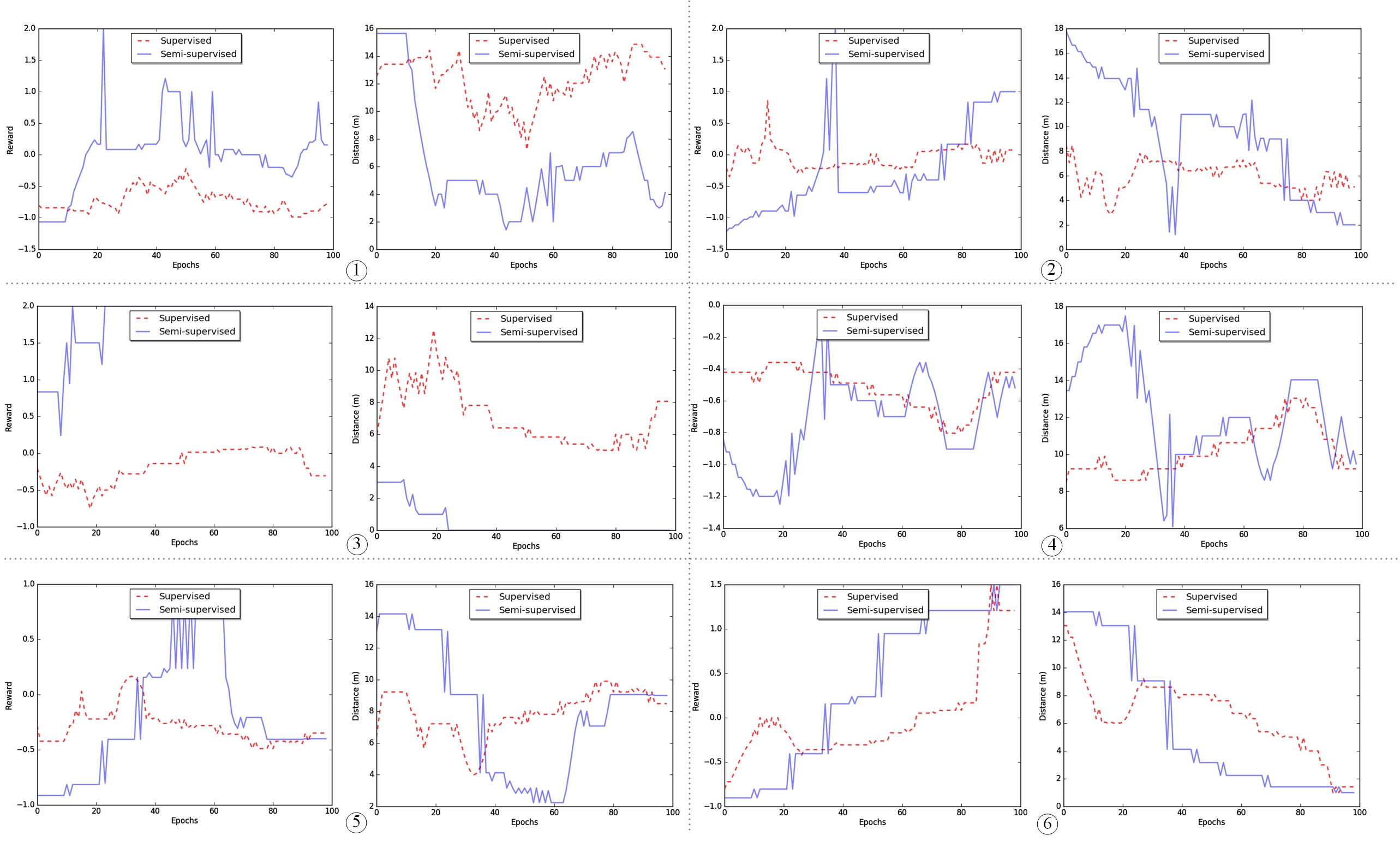}		
	\end{center}	
	\caption{Obtaining rewards and distances in six episodes with a supervised and semi-supervised DRL models.}\label{fig:figset01}	
\end{figure*}

\subsection{Evaluation}	

To implement our proposed semi-supervised DRL model, we adopted the deep reinforcement learning algorithm in which we incorporated a variational autoencoder to generate more rewarding policies and consequently increasing the accuracy of the localization process. The deep neural networks are implemented on Google TensorFlow \cite{abadi2016tensorflow} using the Keras package \cite{chollet2015keras}.

To evaluate the performance of the proposed semi-supervised DRL model, we performed two sets of experiments: one in which the DRL framework uses a fully-connected deep neural network for supervised learning; and the other in which the DRL framework uses a stacked variational autoencoder for semi-supervised learning. 

Figure \ref{fig:figset01} shows the performance of the DRL in terms of the received rewards as well as distance to the true target for both supervised and semi-supervised models in six episodes (c.f. labels 1-6 on the Figure) . In the plots, it can be seen that the agent in the semi-supervised model learns to achieve higher rewards or smaller distances to the target compared to the supervised model.  

Table \ref{tbl:distance_converge} shows that the behavior of the semi-supervised model leads to getting closer to the target points compared to just relying on a supervised model. It also indicates faster steps to reach or get close to the target in the same number of epochs.  
The differences of distances in this table emphasize that the semi-supervised model generates policies that improve the average convergence speed of the localization system by a factor of at least 4.  

\begin{table}
\renewcommand{\arraystretch}{1.2}
\centering
\caption{The average speed of convergence to destination points}
\label{tbl:distance_converge}
\begin{tabular}{l|l|l|l|}
\cline{2-4}
\multirow{2}{*}{}                              & \multicolumn{3}{c|}{Average distance to the target points (meter)} \\ \cline{2-4} 
                                               & as the agent starts    & end of epochs   & difference   \\ \hline
\multicolumn{1}{|l|}{\textbf{Supervised}}      & 9.4        & 7.4      & 2            \\ \hline
\multicolumn{1}{|l|}{\textbf{Semi-supervised}} & 12.8          & 4.3      & 8.5          \\ \hline
\end{tabular}
\end{table}

\begin{figure}
	\begin{center}		
		\includegraphics[width=0.5\textwidth]{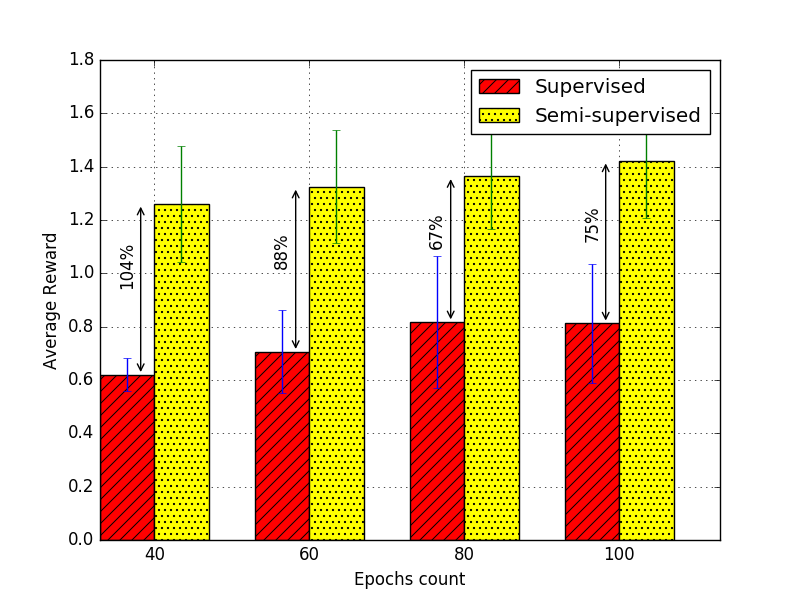}
		
	\end{center}
	
	\caption{The average rewards that are obtained by DRL over different epoch counts using supervised model versus semi-supervised model.}\label{fig:reward_comp}
	
\end{figure}

\begin{figure}
	\begin{center}		
		\includegraphics[width=0.5\textwidth]{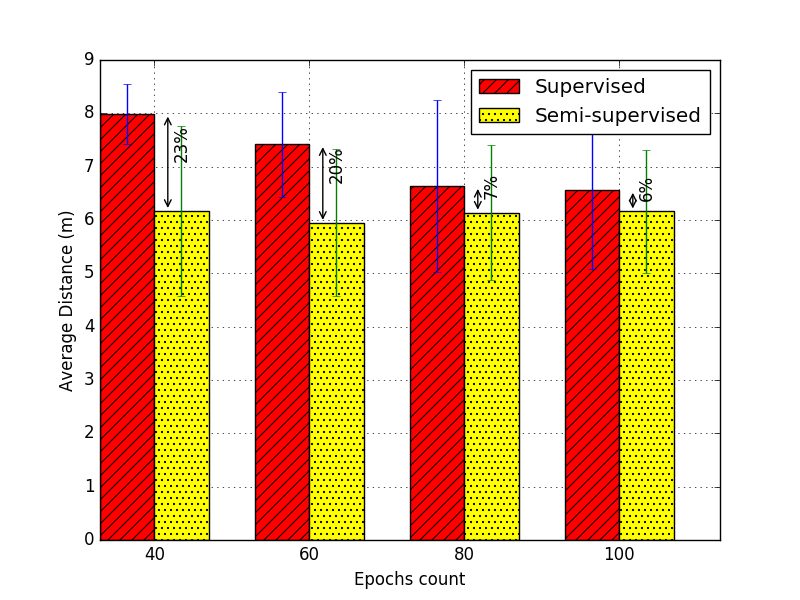}
		
	\end{center}
	
	\caption{The average distance to the target over different epoch counts using supervised model versus semi-supervised model.}\label{fig:dist_comp}
	
\end{figure}

In Figures \ref{fig:reward_comp} and \ref{fig:dist_comp}, the comparison of utilizing the semi-supervised model versus the supervised model along a different number of epochs shows the efficacy of the semi-supervised approach in handling the localization problem. The results in Figure \ref{fig:reward_comp} show that the semi-supervised model reaches a higher reward faster compared to the supervised model while keeping its rewards trend stable. From this figure, it can be seen that the semi-supervised model gains at least 67\% more rewards compared to the supervised model. In addition, the semi-supervised model achieves about twice the rewards of the supervised model. This result can be translated to the original measurement where we want to know the effect of the models on the accuracy of the localization as depicted in Figure \ref{fig:dist_comp}. Figure \ref{fig:dist_comp} shows the average distance to the target points in different number of epochs. Here the semi-supervised model achieves 6\% to 23\% improvement for localization. This result indicates that the unlabeled data helps the VAE to better identify the discriminative boundaries and consequently improves the accuracy of the semi-supervised model. 

\section{Conclusion}\label{sec:Conclusion}

We proposed a semi-supervised deep reinforcement learning framework as a learning mechanism in support of smart IoT services. The proposed model uses a small set of labeled data along with a larger set of unlabeled ones. The current work is the first attempt that extends the semi-supervised reinforcement learning approach using deep reinforcement learning. The proposed model consists of a deep variational autoencoder network that learns the best policies for taking optimal actions by the agent. 

As a use case, we experimented with the proposed model in an indoor localization system. Our experimental results illustrate that the proposed semi-supervised deep reinforcement learning model is able to generalize the positioning policy for configurations where the environment data is a mix of labeled and unlabeled data and achieve better results compared to using a set of only labeled data in a supervised model. The results show an improvement of 23\% on the localization accuracy in the proposed semi-supervised deep reinforcement learning model. Also, in terms of gaining rewards, the semi-supervised model outperforms the supervised model by receiving at least 67\% more rewards.

This study shows that IoT applications in general, and smart city applications in specific where context-awareness is a valuable asset can benefit immensely from unlabeled data to improve the performance and accuracy of their learning agents. Furthermore, the semi-supervised deep reinforcement learning is a good solution for many IoT applications since it requires little supervision by giving a rewarding feedback as it learns the best policy to choose among alternative actions.

\section*{Acknowledgment}
The authors would like to thank Western Michigan University Libraries for providing the experimental testbed and space needed to conduct this research.

\bibliographystyle{IEEEtran}

\bibliography{references}

\end{document}